\documentclass[11pt,twoside]{article}
\usepackage{epsfig}
\newcommand{\mdot}{{\dot{M}}}
\begin{document}

\title{The chemical evolution of the Galaxy: \\ the importance of stars with an initial mass larger than 40M$_{\odot}$}

\author{E. De Donder and D. Vanbeveren}

\maketitle

\begin{abstract}
\noindent In the present paper we investigate in how far stars with an initial mass larger than 40M$_{\odot}$ affect the
chemical enrichment of the Galaxy. We illustrate the importance for chemical yields of a most up to date treatment of the
various stellar wind mass loss episodes in stellar evolutionary codes and we discuss the effects of a possible supernova-like
outburst prior to massive black hole formation.
\end{abstract}

\section{Introduction}

Many existing chemical evolutionary models (CEMs) of galaxies follow in a rather sloppy way the evolution of massive stars with an initial mass larger than 
40M$_{\odot}$ and their contribution to the enrichment of the interstellar matter (ISM). To illustrate, the chemical yields published by 
Woosley $\&$ Weaver (1995) (WW95) are commonly used in CEMs. However, WW95 does not account for the effects of stellar wind (SW) mass loss during different
evolutionary phases of massive single stars and massive binaries (i.e. the OB-type phase, the luminous blue variable (LBV) phase, the red supergiant (RSG)
phase, the Wolf-Rayet (WR) phase and, last not least, the Roche lobe overflow (RLOF) phase). Furthermore, the WW95 tables do not include information on stars
with an initial mass larger than 40M$_{\odot}$. In many CEMs, the effect of this mass range is studied by EXTRAPOLLATING the WW95 results. The authors of these
CEMs argue that the massive star population in the mass range $>$ 40M$_{\odot}$ is much less important than the massive star population in the $\leq$
40M$_{\odot}$ mass range. This is true as far as the number of stars is concerned, but, as we will show in the present paper, this is not true as far as the
chemical enrichment is concerned. An exception on the sloppy treatment discussed above is the work of Portinari et al. (1998) (P98).

In a previous paper (De Donder $\&$ Vanbeveren, 2002a) (DV02) we presented our CEM where we focussed on the effects of binaries, a topic that
was ignored in most of the previously published studies on the chemical evolution of galaxies. In the present paper we will use this CEM in order to illustrate
the importance of massive stars and we will present CEM results which account for the most recent evolutionary computations of massive single and binary stars.
In particular, we will highlight the importance of the $>$ 40M$_{\odot}$ mass range in relation to the recent observations of hypernovae. Since the early
chemical evolution of the Galaxy is governed by massive stars, we will compare our simulations with the observed abundances of very metal-poor stars which are
presumably formed in the early Galaxy. 

In section 2 we describe our CEM. The results of the simulations are discussed in section 3 and compared to the observations.

A similar study on the influence of metallicity dependent SW yields and black hole (BH) formation on galactic chemical evolution has been done by
Prantzos et al. (1994a) and Prantzos (1994b). In section 4 we compare our results with these works.

\section{The chemical evolutionary model}

Our CEM has been described in DV02. We assume that the Galaxy forms in two phases of major infall, during which the halo-thick disk and the thin disk are formed
respectively. For the parameters in this formation model we adopt the values that correspond with the best model A  of Chiappini et al. (1997). We use the
functional star formation formalism of Talbot $\&$ Arnett (1975) (see also Chiosi (1980)) and assume a star formation stop if the surface gas density
is below 7 M$_{\odot}$/pc$^{2}$ (Gratton et al. (1996)).

We have an extended library of evolutionary computations of single stars and interacting binaries that covers the following parameter space: 0.1M$_{\odot}$
$\leq$ single star or primary star mass $\leq$ 120M$_{\odot}$¤, 0 $<$ binary mass ratio $\leq$ 1, 1 day $\leq$ binary orbital period $\leq$ 3650 days, 0.001
$\leq$ metallicity Z $\leq$ 0.02. The evolutionary calculations account for convective core overshooting during the core hydrogen burning phase (CHB) as it was
introduced by Schaller et al. (1992), for the most recent formalisms of SW mass loss during all phases of massive star evolution (see section
2.1), for the process of RLOF and mass transfer in case A and case Br binaries, for the common envelope process in case Bc and
case C binaries, for the common envelope and spiral in process in binaries with a compact companion. All our evolutionary
tracks of massive single stars and binaries are computed till the end of core helium burning (CHeB.). To obtain the supernova
(SN) yields and the mass of the remnant compact star, we link our end-CHeB models with the nucleosynthesis computations of
WW95. Hereby we follow the procedure that is outlined by P98. From the tabulated data of WW95, P98 derived a relation between
the carbon-oxygen (CO) core mass  and the amount of different elements ejected in the explosion of the core. We use this
relation and interpolate with respect to the CO core mass and metallicity to get the ejecta from the CO cores of our computed
models. The remaining layers above the CO core are assumed to be ejected without being modified by the outgoing shock wave.
Remark that the CO core masses of interacting massive binary components are smaller than their single counterparts because of
the enhanced mass loss by RLOF and therefore in general produce lower amounts of metals. On the other hand accretion stars in
binaries may develop larger convective cores and generate higher metal yields than expected from their initial mass at birth.  

The temporal evolution of the stellar population in a galaxy is calculated with a detailed population number synthesis (PNS) code that follows in detail both the
evolution of single and binary stars. For the formation of SNIa's we consider two possible scenarios: the single degenerate scenario (Hachisu et
al. 1999a, 1999b and references therein) and the double degenerate scenario (Iben $\&$ Tutukov (1984); Webbink (1984)). A detailed study on
the temporal evolution of the SNIa and core collapse SN rates in disk galaxies has been presented in De Donder $\&$ Vanbeveren (2002b).  

In DV02 we assumed that all stars with an initial mass $>$ 40M$_{\odot}$ finally collapse into a BH without the ejection of matter. This may not be
correct and we will show in section 3 that alternative assumptions may have a profound effect on the chemical enrichment. If BH formation above 40M$_{\odot}$
is preceded by a SN-like outburst, the SW mass loss during the various evolutionary phases of the BH progenitors determines critically the chemical yields. The
next subsection summarises the SW formalisms that we use.

\subsection{Stellar wind formalisms and their effect on the chemical yields}

Stellar wind mass loss may affect significantly the evolution of the convective core of a massive star and, therefore, SW mass loss can significantly affect
the chemical yields of massive stars. We distinguish four SW phases which are important for massive star evolution: the OB phase including the eventual LBV
phase, the RSG phase and the WR phase. The formalisms that we adopt in our evolutionary code since 1997-98 have been discussed in Vanbeveren et al. (1998a,
1998b, 1998c). To understand their importance for CEMs, the following summary is appropriate.

\subsubsection{Stars with an initial mass $\geq$40M$_{\odot}$}

Based on the observations of LBVs one may suspect that stars with an initial mass $\leq$ 40M$_{\odot}$ experience an LBV phase at the end of CHB and/or hydrogen
shell burning (HSB) which is accompanied by a very violent SW mass loss phase. In the Solar Neighbourhood, the lack of RSGs with an initial mass
$\geq$ 40M$_{\odot}$ (corresponding roughly to stars with M$_{bol}$ $\leq$ -9.5) may be attributed to this process so that a working hypothesis for stellar
evolutionary calculations may be the following:

\begin{itemize}
\item[] the  $\mdot$ during the LBV + RSG phase of a star with an initial mass larger than 40 M$_{\odot}$ must be sufficiently
large to assure a RSG phase which is short enough to explain the lack of observed RSGs with M$_{bol}$ $\leq$ -9.5.
\end{itemize}

Obviously, binary components with an initial mass larger than $\sim$40M$_{\odot}$ will obey this criteria as well which means that the following scenario
applies:  when a primary with an initial mass $\geq$ 40M$_{\odot}$ starts its LBV phase before the RLOF (case B and case C systems), the evolution is governed by
the LBV SW, and the RLOF (and thus mass-transfer) is suppressed. Case A systems in this mass range may evolve more conservatively but we treat them in the same
way as case B systems. Test calculations show that this simplification only marginally affects our CEM simulations. Also the Magellanic Clouds show a deficiency
of RSGs with M$_{bol}$ $\leq$ -9.5 (Humphreys $\&$ McElroy (1984)) so that the single star and binary evolutionary scenario outlined above may apply in low
metallicity regions as well. Notice that the foregoing 'LBV scenario' for massive stars with an initial mass $>$ 40M$_{\odot}$ in general, massive close
binaries with an initial primary mass $>$ 40M$_{\odot}$ in particular was introduced more than a decade ago (Vanbeveren (1991)), and has been used in all
our PNS calculations since 1997. Besides the fact that one has to account for the CNO yields lost by the stars due to the SW mass loss, an important property
of the foregoing 'LBV scenario' is that all stars with an initial mass $>$ 40M$_{\odot}$ are hydrogen deficient during most of their CHeB phase. In the
paragraph on the WR mass loss rates we will explain in how far this affects the CEM results.

In our stellar evolutionary code we apply since 1997-1998 an RSG SW mass loss formalism that is based on the observations of Jura (1987) and
Reid et al. (1990).  More details are given in Vanbeveren et al. (1998a, 1998b, 1998c). The RSG mass loss affects the blue- and redward
evolution in the HR-diagram during the HSB-CHeB burning phases of single stars with an initial mass 20 $\leq$ M/M$_{\odot}$
$<$ 40. The RSG SW removes the hydrogen rich layers and it determines the moment that the massive star becomes a hydrogen deficient CHeB star,
resembling a WR star.  We use stellar evolutionary calculations with a  $\sqrt{Z}$-dependent  $\mdot$-formalism for RSGs. This
means that for the early evolution of the Galaxy, the effect of RSG mass loss on the overall chemical yields of massive stars
is small.

\subsubsection{WR SW mass loss}

Using a hydrodynamic atmosphere code where the SW is assumed to be homogeneous, Hamann $\&$ Koesterke (1998) determined 
$\mdot$-values for a large number of WR stars.  Since then evidence has grown that these winds are clumpy and that a
homogeneous model overestimates $\mdot$  with typically a factor 2-4 (Hillier et al. (1996); Moffat et al. (1996); Schmutz
(1997); Hamann $\&$ Koesterke (1998)).  In the period 1997-1998 massive single star and binary evolutionary calculations were
published Vanbeveren et al. (1998a, 1998b, 1998c) for which we adopted the lower WR mass loss rates as inferred from the
studies listed above. We assumed a (simple) relation between  $\mdot$ ( in M$_{\odot}$/yr) and the stellar luminosity L (in
L$_{\odot}$) of the form  $\log$($\mdot$)=a$\log$L+b, and we tried to find appropriate values for the constants a and b by
accounting for the following criteria and observations known at that time:

\begin{itemize}
\item  the WN5 star HD 50896 (WR 6) has a luminosity $\log$ L = 5.6-5.7 and a $\log$($-\mdot$) = -4.4$\pm$0.15 (Schmutz (1997))

\item  the $\log$(-$\mdot$) of the WNE component of the binary V444 Cyg (WR 139) derived from the observed orbital period variation is $\sim$-5
(Khaliullin et al. (1984); Underhill et al. (1990)). Its orbital mass is 9M$_{\odot}$ and using a mass-luminosity relation holding for
WNE-binary components (Vanbeveren $\&$ Packet (1979); Langer (1989)) it follows that its $\log$L=5

\item  the observed masses of BH components in X-ray binaries indicate that stars with initial mass $>$ 40M$_{\odot}$ should end their life with a mass larger
than 10M$_{\odot}$ (= the mass of the star at the end of CHeB)

\item the WN/WC number ratio predicted by stellar evolution depends on the WR  $\mdot$-formalism. Therefore, last but not
least, we looked for a and b values which predict the observed WN/WC number ratio ($\approx$1) for the solar neighbourhood.
\end{itemize}

This exercise allowed us to propose the following relation:

\begin{equation}
\log(-\mdot) = \log L -10
\end{equation}
                    			     
Since then more WR stars have been investigated with detailed atmosphere codes including the effects of clumping. They are listed in table
1. Figure 1  illustrates that relation 1 still fits fairly well these new observations. Interestingly,
the WC6 star OB10-WR1 in the association OB10 of M31 has been investigated by Smartt et al. (2001) and also fits relation
1 [the star is plotted in figure 1 as well].

Nugis $\&$ Lamers (2000) (NL) proposed an alternative WR mass loss relation based on a large sample of Galactic WR stars. In
table 1 the predictions from equation 1  and from the NL-formula are compared together with the
spectroscopic results. As can be noticed, NL is not more accurate than our relation. 

Remark that NL uses very uncertain WR-calibrations of distance and bolometric correction. Moreover, some of the WR stars listed by NL have a luminosity and mass
loss rate which differs significantly from the results of detailed non-LTE line blanketed spectroscopic analyses
(table 1). We have the impression that at least for the WNE stars, the luminosity in the NL-list is rather large
(see also Hamann $\&$ Koesterke (2000)).  

The Z-dependency of WR mass loss rates deserves some attention. Nugis and Lamers adopt a  $\mdot$-relation which is
Z-dependent where, as usual, Z is defined as the abundance of all elements heavier than helium. This means that they
intrinsically assume that WN and WC stars with the same luminosity and belonging to the same stellar environment (thus having
the same iron abundance) have quite a different mass loss rate. However, there is no observational evidence for this to be
true. Even more, when the SW is radiation driven, one expects that the heavy elements (primarily iron) are the main drivers and
that the $\mdot$ depends mainly on the iron abundance X$_{Fe}$. Our evolutionary library contains calculations with and
without a  $\sqrt{X_{Fe}}$-dependency of the $\mdot$  during the hydrogen deficient CHeB phase of a massive star.

\begin{table}[h] 
\centering

\begin{tabular*}{90mm}{cccccc}\hline

 WR number & log L & log(-$\mdot$) & DVB & NL & ref. \cr \hline  
 WR 6 & 5.45 & -4.4 & -4.5 & -4.8 & (1) \cr
{ WR 147}&{ 5.65}&{ -4.6}&{ -4.3}&{ -4.4}&{ (2)}\cr
{ WR 111}&{ 5.3}&{ -4.8}&{ -4.7}&{ -4.9}&{ (3)}\cr
{ WR 90}&{ 5.5}&{ -4.6}&{ -4.5}&{ -4.7}&{ (4)}\cr
{ WR 135}&{ 5.2}&{ -4.9}&{ -4.8}&{ -4.8}&{ (4)}\cr
{ WR 146}&{ 5.7}&{ -4.5}&{ -4.3}&{ -4.4}&{ (4)}\cr
{ WR 11}&{ 5}&{ -5.1}&{ -5}&{ -5}&{ (5)}\cr
{ WR 123}&{ 5.7}&{ -4.14}&{ -4.3}&{ -4.5}&{ (6)}\cr
{ WR 139}&{ 5}&{ -5}&{ -5}&{ -4.5}&{ (7)}\cr\hline
\end{tabular*}
\caption{The luminosity and the mass loss rates of WR stars determined with NLTE atmosphere
models and assuming non-homogeneous stellar winds. We compare with values predicted by equation 1 and with the
formula proposed by NL. Ref. (1) = Schmutz (1997), (2) = Morris et al. (2000), (3) = Hillier $\&$ Miller (1998), (4) = Dessart
et al. (2000), (5) = Nugis et al. (1998), (7) =Underhill et al. (1990)}

\end{table}

\begin{figure}[h]
\centering
\epsfig{file=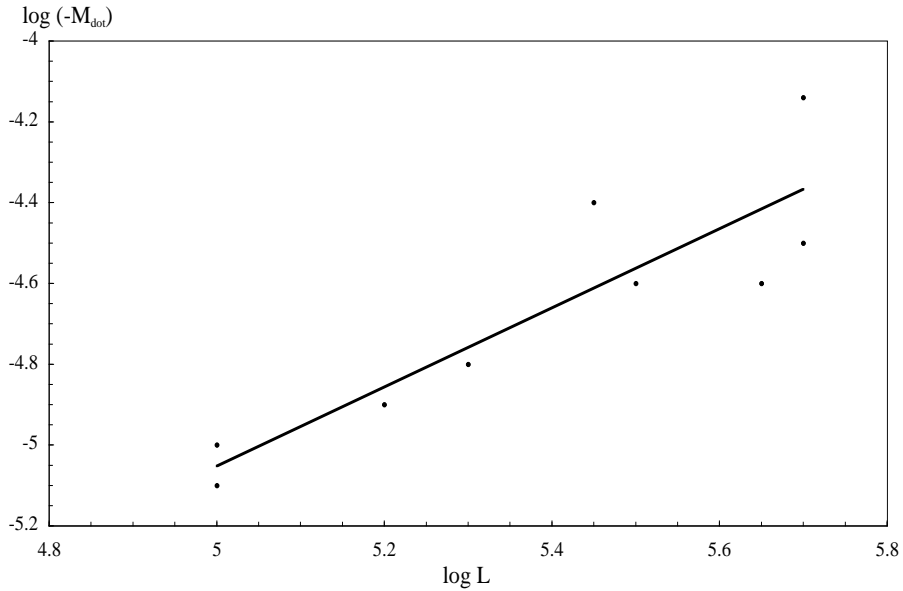, height=8cm,width=12cm}
\caption{The log(-$\mdot$) vs. log L plot for WR stars.}

\end{figure}

\subsubsection{Evolutionary results}

In 1997-1998 we calculated evolutionary tracks of massive single stars and binary components adopting the RSG mass loss formalism for single stars discussed
above and equation 1 with and without a heavy element dependency for the SW mass loss rate at the moment that the
star becomes a hydrogen deficient CHeB object. These tracks are part of our library. All our results published since 1997-1998
(see the papers cited in section 2) rely on this library and thus rely on equation 1. Of particular importance for
the SN rates and BH formation are the pre-SN evolutionary masses (=M$_{f}$) predicted by our calculations. They are shown in
figure 2. We notice here that:

\begin{itemize}

\item Galactic massive stars have a pre-SN mass between 2M$_{\odot}$ and (20-25)M$_{\odot}$ which allows to explain in a straightforward way the observed BH
masses in the standard high mass X-ray binary Cyg X-1 and in a number of low mass X-ray binaries.

\item If the WR (and/or RSG) stellar wind mass loss rate depends on the heavy metal abundance as described earlier, the pre-SN masses of massive stars may be
significantly larger at low metallicity.

\end{itemize}

For the CEMs the effect of RSG and WR mass loss on the evolution of the CO cores in massive stars is critical. 
When the WR mass loss is small (which is the case in regions with a small iron abundance and when the WR mass loss rates are iron dependent) or when the RSG
mass loss is too small that a massive single star never becomes a WR star (which is the case for all stars with an initial mass $\leq$ 20M$_{\odot}$ and for
stars with an initial mass $\leq$ 40M$_{\odot}$ in regions with a small iron content) the CO core increases in mass during the whole CHeB. At the onset of
carbon burning, the layers outside the carbon burning core mainly consist of oxygen (and, obviously, the layers which were outside the previous He burning core
consist mainly of helium). When at the end of the stars' life, the formation of a compact star (a neutron star (NS) or a BH) is accompanied by matter ejection,
the ISM may be significantly enriched in oxygen (and little carbon).  

However, when the WR mass loss rates are large (which is the case when equation 1 applies) and the massive star
becomes WR like soon after the onset of CHeB, after an initial rapid increase, the CO core decreases in mass leaving behind
layers which may contain a significant amount of carbon. When these layers reach the stellar surface (forming a WC type star)
large amounts of carbon are carried away by further SW mass loss. If at the end of the starsÕ life, the formation of a compact
star is accompanied by matter ejection, also this matter will contain a significant amount of carbon. 

This means that it can be expected that the simulation of the enrichment of carbon and oxygen in the early evolutionary phases of the Galaxy may depend on
whether or not one uses massive star evolutionary results where the WR mass loss rates are iron dependent or not (see section 3).

\begin{figure}[h]
\centering
\epsfig{file=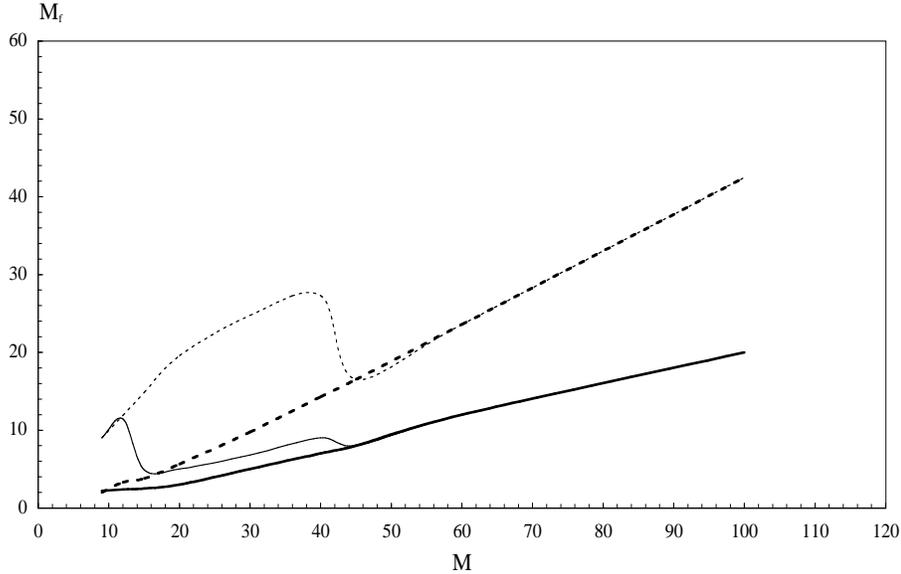,height=8cm,width=12cm}
\caption{The pre-SN masses (M$_{f}$) of massive single (thin lines) and interacting primary (thick lines) stars as predicted by our calculations. The
computations are made for Z=0.02 (full lines) and Z=0.002 (dotted lines).}

\end{figure}

\subsection{Black hole formation}

From the evolutionary calculations of massive stars with the WR mass loss rates discussed in the previous section we conclude that all stars with an initial 
mass $>$ 40M$_{\odot}$ may form massive BHs. Crucial for CEMs is then the question whether or not BH formation in this mass range is preceded by a SN-like
outburst.  Israelian (1999) presented observational evidence that the formation of the BH in the X-ray binary GRO J1655-40 was preceded by the ejection of heavy
elements. The high mass X-ray binary Cyg X-1 contains a massive BH with a mass $\geq$ 10M$_{\odot}$. It is a runaway which may be an indication of the occurrence
of a previous SN explosion (Nelemans et al. (1999)). The latter authors reached similar conclusions when considering the runaway velocity of a number of low
mass X-ray binaries with a massive BH candidate component. The last few years a class of extremely energetic supernova has been recognised (Nomoto
et al. (2001)) and it is tempting to associate them with the formation of a BH (Woosley (1993); Paczinsky (1998); Iwamoto et al. (1998)). The
amount of ejected $^{56}$Ni that is needed to explain the lightcurve of these hypernovae is very large, as large as 0.4-0.7M$_{\odot}$
(Sollerman et al. (2000)). This means that if the formation of all BHs is preceded by a hypernova, their contribution to the iron enrichment of the ISM
during the early phases of the Galaxy could be substantial.

\section{Chemical evolutionary simulation of the early Galaxy}

From the discussion in the previous section it can be expected that the chemical evolutionary simulation of the early Galaxy may depend significantly on the
evolutionary scenario adopted for stars with an initial mass $>$ 40M$_{\odot}$, in particular on the WR type SW mass loss formalism and on the physics of BH
formation. We focus on the elements carbon, oxygen and iron. Using our CEM that includes binary evolution as described in DV02, we explore the effects of the
latter stars by adopting the following stellar evolutionary scenarioÕs:

\underline{Sim1:} all stars with an initial mass $>$ 40M$_{\odot}$ form massive BHs without the ejection of matter; the WR SW mass loss rates obey
equation 1 and they do not depend on the heavy metal content of the gas out of which the stars are formed
 
\underline{Sim2:} all stars with an initial mass $>$ 40M$_{\odot}$ form massive BHs without the ejection of matter; the WR SW mass loss rates obey equation
 1 and are scaled proportional to $\sqrt{X_{Fe}}$  

\underline{Sim3:} all stars with an initial mass $>$ 40M$_{\odot}$ form massive BHs with the ejection of 0.7M$_{\odot}$ of $^{56}$Ni (i.e. the mass of
the BH = the mass of the Fe-Ni core Ð0.7M$_{\odot}$), i.e. we explore the consequences on CEMs when all BHs are accompanied by a hypernova explosion; the WR SW
mass loss rates obey equation 1 and are scaled proportional to  $\sqrt{X_{Fe}}$

\underline{Sim4:} the WR SW mass loss rates obey equation 1 and are scaled proportional to $\sqrt{X_{Fe}}$, all
stars with an initial mass $>$ 40M$_{\odot}$ form massive BHs with the ejection of matter that contains no $^{56}$Ni ; the
amount of ejected matter obviously depends on the physics of BH formation. We explore the consequences when the total amount of
ejected material is chosen so that it contains 4 M$_{\odot}$ of oxygen (respectively 8M$_{\odot}$ of oxygen). 

Figures 3 and 4  show the predicted temporal evolution of the abundance ratios [O/Fe] and [C/Fe] (for
the solar values we use the observed abundances of Anders $\&$ Grevesse (1989) for the Galaxy. To describe the formation of the
Galaxy we adopt the parameters of model A in Chiappini et al. (1997). In the upper initial mass range (30 $\leq$ M/M$_{\odot}$
$\leq$ 40) WW95 calculated the SN yields for several models that differ  in explosion energy. For the computation of our SN
yields (according to the method explained in section 2) we use the results of their model B  but reduce the iron yields with a
factor of two. A reduction of the WW95 iron yields has been shown to give a better agreement with most of the observed
evolution of the abundances (e.g. Timmes et al. (1995)), corresponds with the observational estimates of the amount of iron
synthesized in SN 1987A and SN 1993J (Thomas et al. (1998)) and is within the theoretical uncertainty. Furtheron, our CEM
accounts for the evolution of binaries for which we adopt the standard set of binary parameter values (see DV02) and a binary
frequency of 70$\%$ over the whole mass range independent of time and metallicity. The observational data sources are given in
the legend of the figures.  The observational data of [O/Fe] deserves some attention. The available data sets can be separated
into two groups: the first group consists of the set of Gratton et al. (1996) and the set of Nissen et al. (2002) which predict
[O/Fe] = 0.5-0.7 at [Fe/H] = -2.5 with a tendency of a plateau between [Fe/H] = -2 and [Fe/H] = -1, and a second group which
consists of the observations of Israelian et al. (1998) and Boesgaard et al. (1999) showing that oxygen is overabundant in the
galactic halo with an increase of [O/Fe] from 0.6 to 1 going from [Fe/H] = -1 to Ð3. From a comparison between the simulations
and the observations we draw the following conclusions. 

\begin{itemize}

\item \textit{Our theoretical simulations illustrate the large effect that stars with an initial mass $>$ 40M$_{\odot}$ have on the chemical evolution of a
galaxy. Therefore it is of particular importance to implement the most up to date stellar wind mass loss rate formalisms during the various evolutionary phases
for the considered mass range.}

\item \textit{The observations of [O/Fe] and [C/Fe] at small values of [Fe/H] are not reproduced when a CEM is adopted in which a majority of BHs is formed in a
hypernova explosion with the ejection of a significant  amount of $^{56}$Ni.}

\item \textit{A CEM that assumes a metallicity independent WR SW predicts a significant overenrichment in carbon during the early evolutionary phases of the
Galaxy. We consider this as an indirect evidence that these winds depend on the heavy element abundance, as predicted by the radiation driven stellar wind
theory.}

\item \textit{The evolution of stars with an initial mass $>$ 40M$_{\odot}$ in general, the detailed physiscs of BH formation and the preceding amount of matter
ejection in particular, decide upon the real carbon, oxygen and iron enrichment of a galaxy.}

\end{itemize}

\begin{figure}[h]
\centering
\epsfig{file=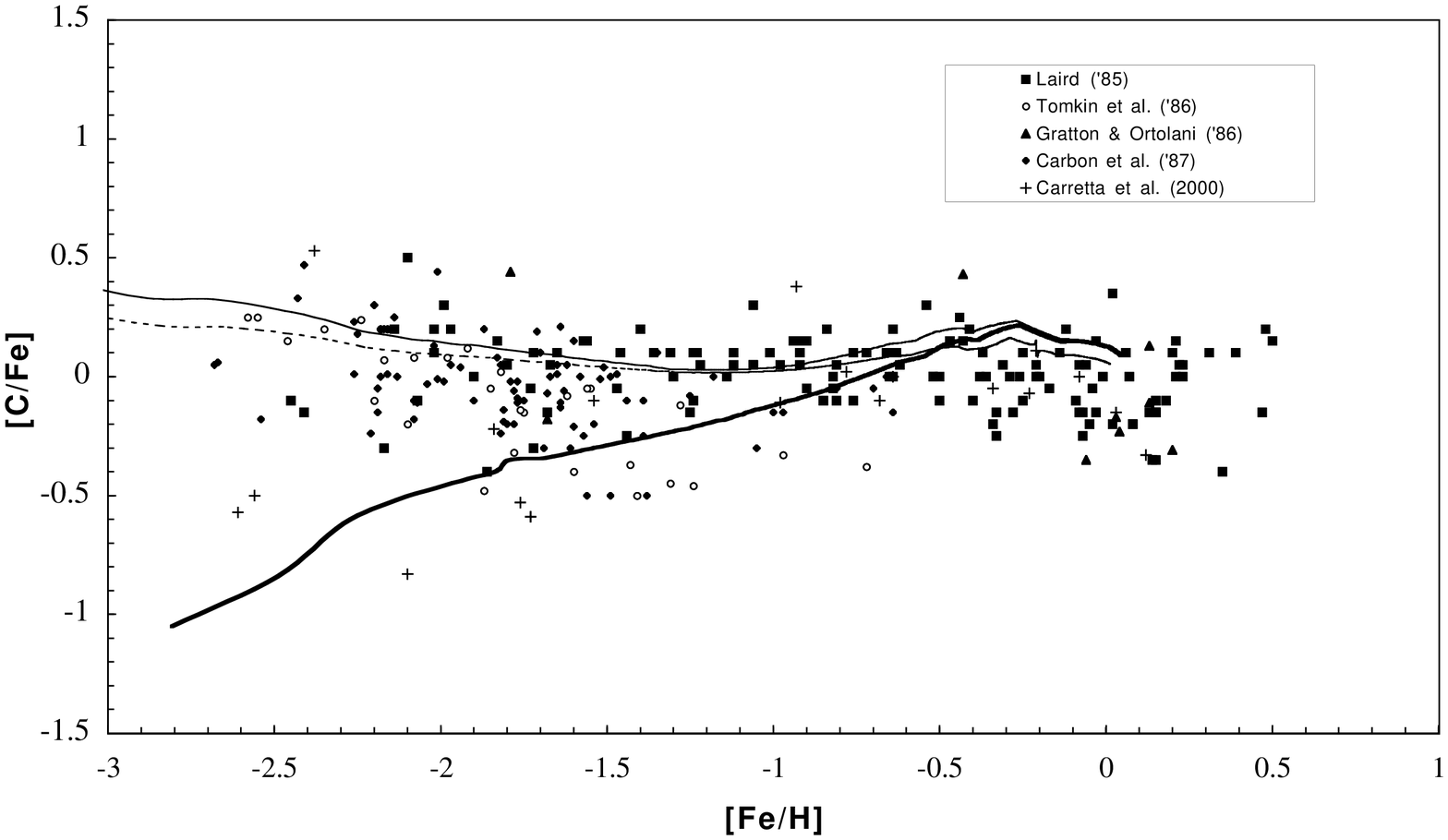,height=8cm,width=12cm}
\epsfig{file=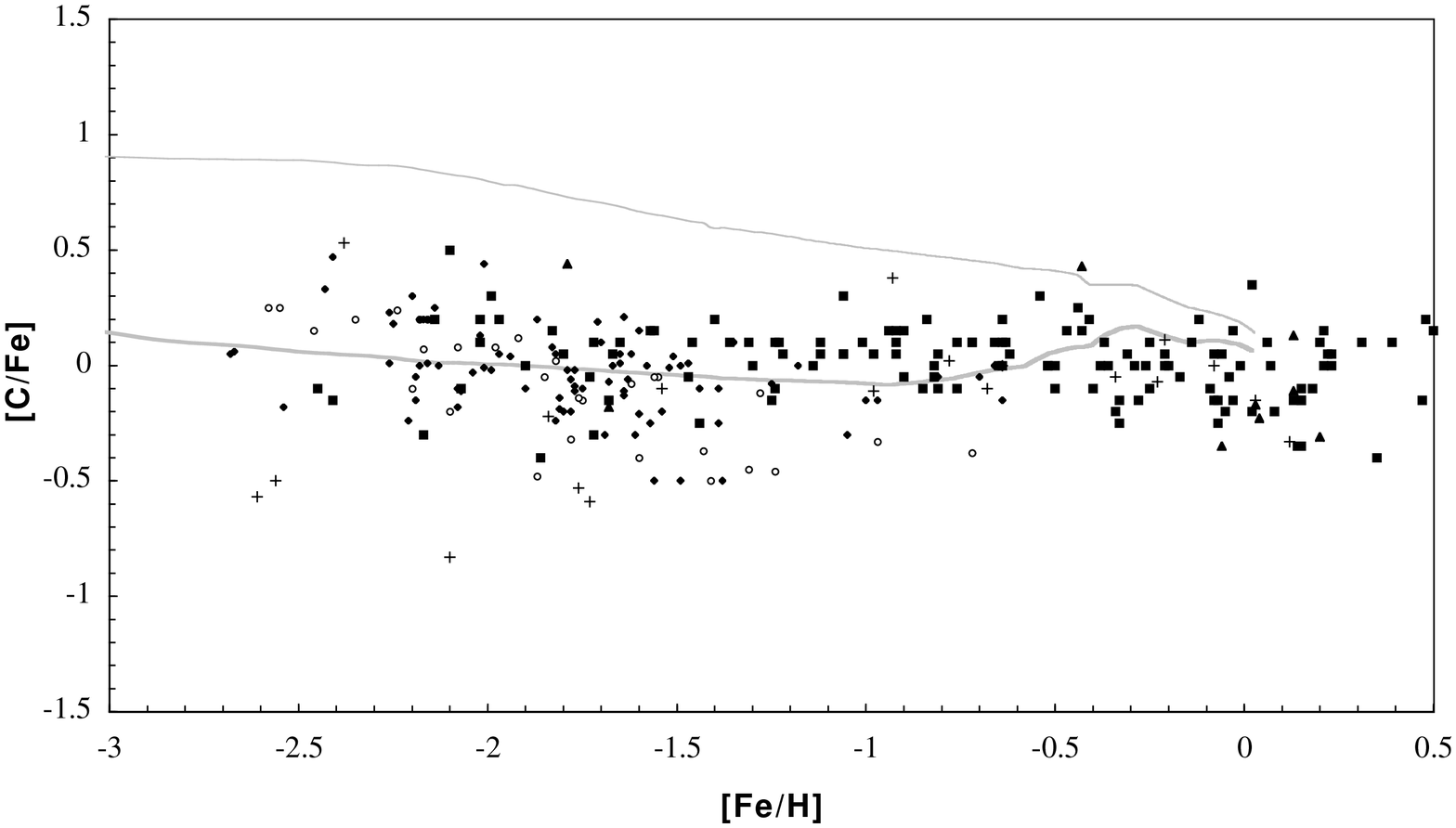,height=8cm,width=12cm}
\caption{The predicted evolutionary behaviour of [C/Fe] as function of time for the different simulations: Sim1 = thin gray line, Sim2 = thick gray line, Sim3 =
thick black line and Sim4 = thin black line.}

\end{figure}

\begin{figure}[h]
\centering
\epsfig{file=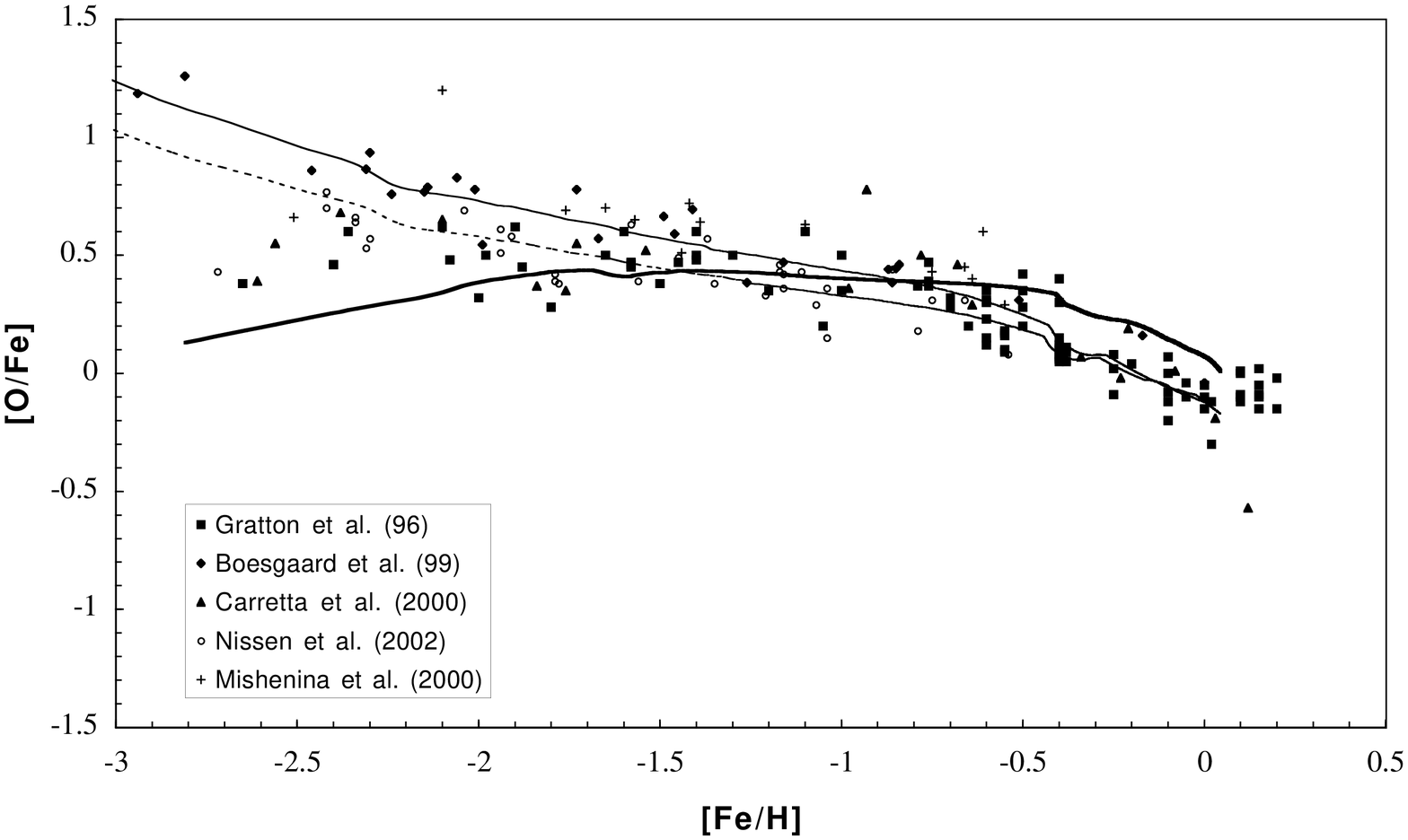,height=8cm,width=12cm}
\epsfig{file=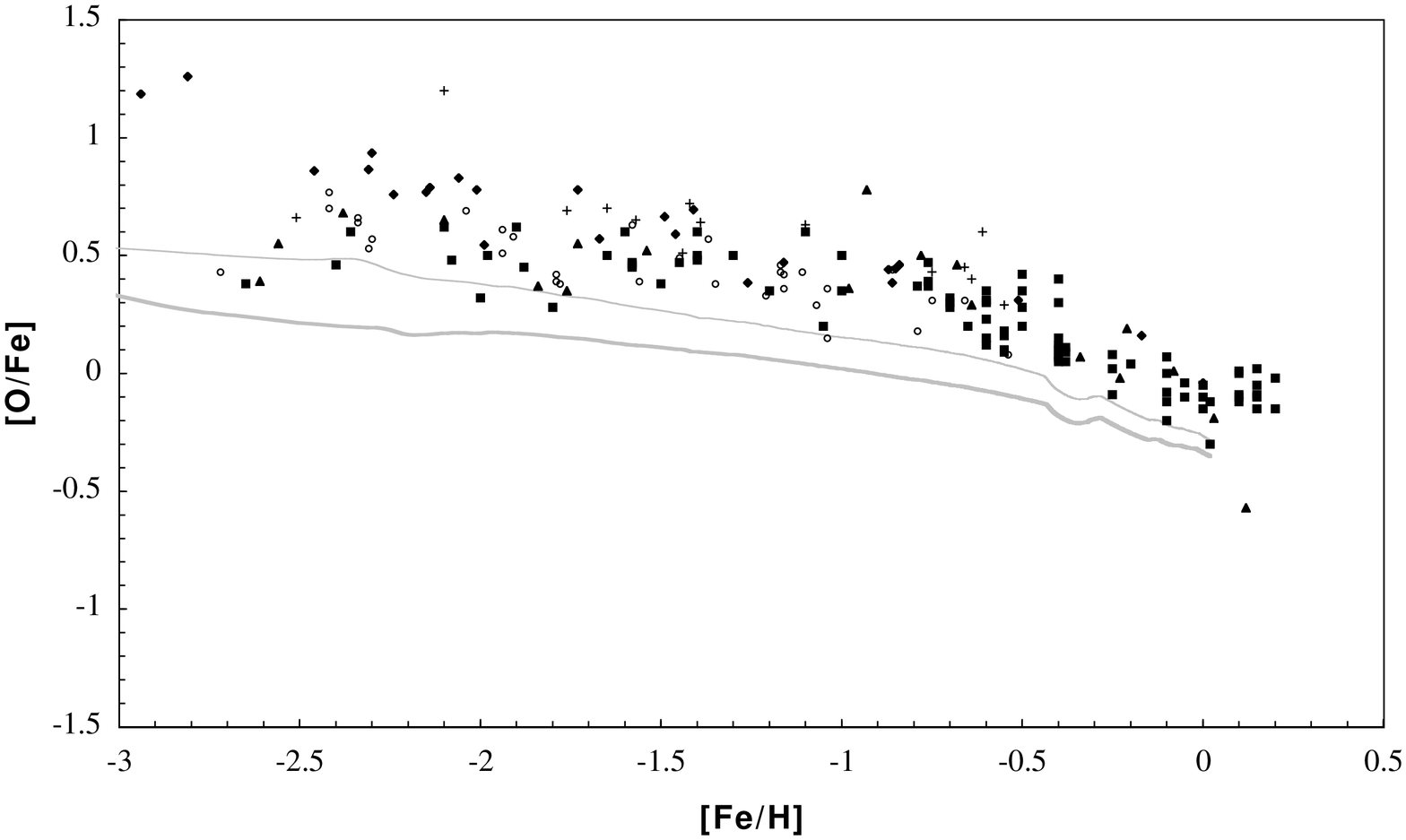,height=8cm,width=12cm}
\caption{The predicted evolutionary behaviour of [O/Fe] as function of time for the different simulations: Sim1 = thin gray line, Sim2 = thick gray line, Sim3 =
thick black line, Sim4 with the ejection of 4M$_{\odot}$  resp. 8M$_{\odot}$ of oxygen = dashed thin black line resp. thin black line.}

\end{figure}

\section{ Comparison with previous work}

Metallicity dependent stellar yields from massive stars with inclusion of SW mass loss have also been computed by Maeder (1992) (M92) and P98. These yields
were respectively implemented into  a CEM by Prantzos et al. (1994a) and by P98 $\&$ Gustafsson et al. (1999) in order to study their influence  on the
evolutionary behaviour of carbon and oxygen in the Galaxy. They all  concluded that the [C/O] vs. [Fe/H] relation for disk stars can be well explained by the
metallicity-dependent behaviour of radiatively driven SWs of massive stars and that there is no real need for a moderate carbon contribution from low and
intermediate mass stars. This importance of metallicity dependent SW yields corresponds with our results though it should be remarked that both M92 and P98 use
mass loss rate prescriptions in their stellar evolutionary computations that are out of date and therefore should not be used in current CEMs. For the WR SW
mass loss they use the Langer formalism (Langer (1989)) which has been proven to overestimate largely  (up to a factor of 4) the mass loss  rate.  Furtheron
they assume a metallicity dependent SW only during the pre-WR phase which makes that at Z $\leq$ 0.002 only the most massive stars ($>$80M$_{\odot}$) may enter
the WR phase and that He, C $\&$ O  enrichment by SW mass loss becomes only important at high metallicity. In our evolutionary model all stars initially more
massive than 40M$_{\odot}$ become WR stars independent of the initial metallicity as it follows from the LBV scenario that we apply for all Z. This assumption
has been invoked to explain the lack of observed RSGs with M$_{bol}$ $\leq$ -9.5 in the MCs and Galaxy.   Aditionally, M92 also concluded that direct BH
formation should already start at
$\sim$(25-30)M$_{\odot}$ to avoid an overproduction of oxygen and to obtain a large helium to metallicity enrichment ratio $\Delta$Y/$\Delta$Z. Extrapolating
this result to low metallicities would imply that stars above 40M$_{\odot}$ are totally irrelevant for the chemical evolution during the early Galaxy contrary
to our results. However M92 did not do the complete galactic evolution. Prantzos (1994b) reanalysed the study of M92 and came to a major conclusion that the
observations are better fitted when all massive stars up to 100M$_{\odot}$ contribute, via SWs and/or SNe, to the chemical enrichment of the Galaxy  which
agrees with our conclusions.

\end{document}